\documentclass[pre,superscriptaddress,aps, twocolumn, showpacs, longbibliography, footnoteinbib]{revtex4-2}
\usepackage [utf8]{inputenc}

\usepackage{graphicx}
\usepackage{amssymb}   
\usepackage{amsfonts}
\usepackage{amsmath}
\usepackage{dsfont}
\usepackage{natbib}
\usepackage{dcolumn}   
\usepackage{bm}        
\usepackage[mathscr]{eucal}
\usepackage[dvipsnames]{xcolor}
\usepackage[colorlinks, linkcolor=OrangeRed,citecolor=RoyalBlue,urlcolor=NavyBlue]{hyperref}
\usepackage[all]{hypcap} 
\usepackage{mathtools}
\usepackage{dsfont}

\definecolor{myblue}{rgb}{0,0,0.75}

\newcommand{\ep}{\varepsilon}

\newcommand{\bra}[1]{\ensuremath{\left\langle#1\right|}}
\newcommand{\ket}[1]{\ensuremath{\left|#1\right\rangle}}
\newcommand{\brakett}[2]{\ensuremath{\left\langle {#1} | {#2} \right\rangle}}

\newcommand\mean[1]{\ensuremath{\left\langle#1\right\rangle}}

\newcommand\lrp[1]{\left(#1\right)}
\newcommand\lrb[1]{\left[#1\right]}

\newcommand\abs[1]{\left|#1\right|}

\newcommand\caseS[1]{\left\{#1\right.}
\newcommand{\lra}{\quad \Leftrightarrow \quad}

\newcommand{\be}{\begin{equation}}
\newcommand{\ee}{\end{equation}}
\def\ba{\begin{aligned}}
\def\ea{\end{aligned}}
\newcommand{\bea}{\begin{eqnarray}}
\newcommand{\eea}{\end{eqnarray}}
\def\bes{\begin{subequations}}
\def\ees{\end{subequations}}
\def\bal{\begin{align}}
\def\eal{\end{align}}

\renewcommand{\Re}{{\rm \, Re\,}}
\renewcommand{\Im}{{\rm \, Im\,}}

\newcommand{\tr}{{\rm Tr}}
\newcommand{\G}{\mathcal{G}}
\newcommand{\s}{\mathcal{S}}
\newcommand{\h}{\mathcal{H}}
\newcommand{\z}{\mathcal{E}}

\newcommand{\Lk}[1]{\ket{L_{#1}}}  
\newcommand{\Lb}[1]{\bra{L_{#1}}}
\newcommand{\Rk}[1]{\ket{R_{#1}}}
\newcommand{\Rb}[1]{\bra{R_{#1}}}
\newcommand{\LRbk}[2]{\brakett{L_{#1}}{R_{#2}}}

\newcommand{\rev}[1]{{\color{black}#1}}
\newcommand{\revt}[1]{{\color{black}#1}}

\usepackage{soul}
\usepackage{ulem} 

\usepackage{graphicx}
\usepackage{amsmath,amssymb,bm}
\usepackage{enumerate}
\usepackage{braket}        

\graphicspath{{./figures/}}

\begin{document}

\title{Non-Hermitian Rosenzweig-Porter random-matrix ensemble: Obstruction to the fractal phase}
\author{Giuseppe De Tomasi}
\affiliation{Department of Physics, University of Illinois at Urbana-Champaign, Urbana, Illinois 61801-3080, USA}
\author{Ivan M. Khaymovich}
\affiliation{Max Planck Institute for the Physics of Complex Systems, N\"othnitzer Stra{\ss}e~38, 01187-Dresden, Germany}
\affiliation{Institute for Physics of Microstructures, Russian Academy of Sciences, 603950 Nizhny Novgorod, GSP-105, Russia}
\affiliation{Nordita, Stockholm University and KTH Royal Institute of Technology Hannes Alfv\'ens v\"ag 12, SE-106 91 Stockholm, Sweden}

\begin{abstract}
We study the stability of non-ergodic but extended (NEE) phases in non-Hermitian systems. For this purpose, we generalize the so-called Rosenzweig-Porter random-matrix ensemble (RP), known to carry a NEE phase along with the Anderson localized and ergodic ones, to the non-Hermitian case.
We analyze, both analytically and numerically, the spectral and multifractal properties of the non-Hermitian case. We show that the ergodic and the localized phases are stable against the non-Hermitian nature of matrix entries. However, the stability of the fractal phase depends on the choice of the diagonal elements. For purely real or imaginary diagonal potential the fractal phase is intact, while for a generic complex diagonal potential the fractal phase disappears, giving the way to a localized one.
\end{abstract}
\maketitle

\section{Introduction}
The study of non-Hermitian many-body systems has emerged as a new paradigm to describe open or dissipative systems with gain and loss. Non-Hermitian systems uncover a rich phenomenology, describing unique effects, e.g., non-Hermitian skin effect, generalized topological phases, measurement-induced phase transitions~\cite{Slager2020boundary,Okuma20,Yao_2018,Rudner_2009,Hu_2011,Esaki_2011,Gong_2018,Schomerus_13,Ashida2020,moiseyev_2011,Skinner_2019,Zabalo_2022,Potter_2021,
Feinberg_1999,Molinari_2009,Huang_2020,Ryu_2021}
which are not possible in Hermitian systems. This growing interest in non-Hermitian systems is also motivated by
advancement in controlled experimental techniques, which allow probing non-Hermitian systems in several contexts, ranging from photonic~\cite{Gonzalo2013quantum,Maximo2015spatial,celardo2017localization,Cottier2019microscopic,Maximo2019Anderson} to topological phases of matter~\cite{Ashida2020,Guo2009,Ruter2010,Bertoldi2017,Esslinger2019,Zhen2015,Weimann2017,RevModPhys_Flore_21}.

Anderson localization (AL) is a milestone of condensed matter physics and describes the localization of non-interacting particles subjects to strong quench disorder~\cite{Anderson1958,Evers2008Anderson}. AL has been extended and extensively studied in non-Hermitian systems\rev{, both in terms of eigenstates~\cite{Hatano_96} and eigenvalue statistics (see,e.g.,~\cite{Shukla_2001_non-Herm,Garcia_2002_non-Herm,Bohigas_2013_non-Herm})}. These studies pointed out that non-Hermitian terms favor delocalization in the systems, as they suppress interference effects, which are crucial for AL. For instance, Hatano and Nelson in \rev{their seminar work} Ref.~\cite{Hatano_96} showed that even in one dimension, where all the eigenstates exponentially localize for any amount of disorder, undergoes a metal-insulator transition if the system is subject to an imaginary vector potential.
%
\begin{figure}[ht!]
\label{fig:Picture}
    \includegraphics[width=1.\columnwidth]{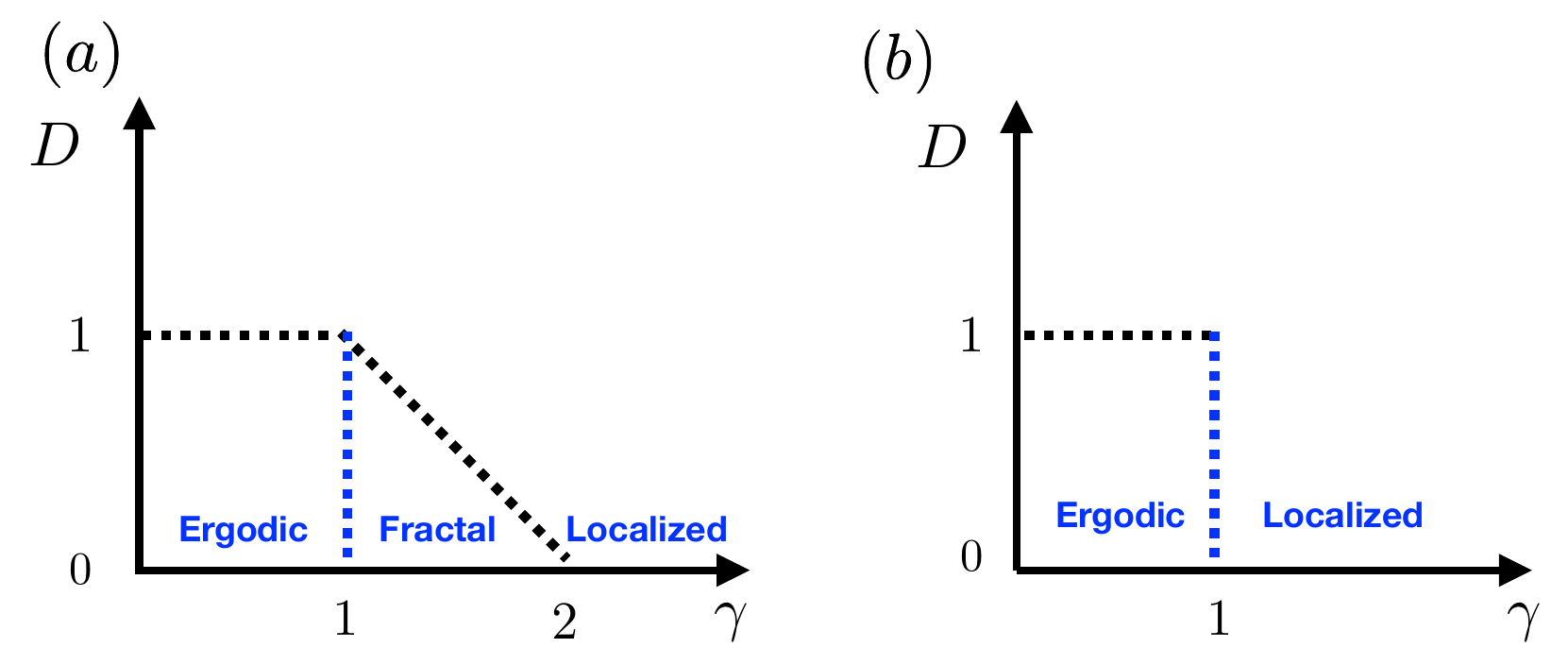}
    \caption{\textbf{Phase diagram of the Rosenzweig-Porter model} $\gamma$ is the parameter of the RP model, tuning the disorder strength. $D$ is the fractal dimension of the wave function, the phase is ergodic for $D=1$, localized for $D=0$, while for $0<D<1$ it is fractal.
    (a)~phase diagram of the Hermitian and non-Hermitian with purely real/complex potential RP model. (b)~phase diagram for the general non-Hermitian RP case.
    }
\end{figure}
Furthermore, the stability of disorder-induced localization in interacting quantum systems~\cite{Basko06,Pal2010, oganesyan2007localization,gornyi2005interacting} has been recently investigated with respect to non-Hermiticity~\cite{Ueda_2019} and shown to be stable. The generalization of AL to the interacting case, dubbed as many-body localization (MBL)~\cite{Nandkishore_2015,Abanin_2019,ALET2018498}, has emerged as the paradigm of ergodicity-breaking in quantum many-body systems. At strong disorder, the system is in the MBL phase and described by an extensive number of conserved quantities, revealing a robust form of emergent integrability~\cite{huse2014phenomenology, ROS_2015,Imbrie2016}.
Instead, in an ergodic phase, the system is delocalized and thermal, meaning that at a long-time, the evolution of local observable is well described by thermodynamic ensembles~\cite{Polkovnikov_2011, DAlessio2016ETH}.

The characterization of disorder quantum systems and their delocalization-localization transition is a challenging task. Random-matrix models have been often used to overcome some of these difficulties. For example, a so-called power-law random banded matrix ensemble~\cite{PLRBM,Evers2008Anderson} shares many aspects with the
single-particle Anderson localization problem on a 3D lattice. Both models have a delocalization-localization transition at finite disorder strength and an emergent multifractality at the critical point.

With the advent of MBL, the study of non-ergodic extended (NEE) phases, e.g., phases composed of multifractal states, has become relevant.
Indeed, MBL being localized in the coordinate space of an interacting system, corresponds to the ergodicity-breaking transition in the Fock/Hilbert space with multifractal states for the entire MBL phase~\cite{De_Luca_2013,Luitz15,Mace_Laflorencie2019_XXZ,QIsing_2021}.
This brings the necessity to understand NEE phases, and to find analytically tractable random matrix models, which host NEE phases. Furthermore, a better characterization of NEE phases of matter could shed a light on optimization algorithms, such as quantum annealing~\cite{smelyanskiy2020non,kechedzhi2018efficient}, or on a non-ergodic phase of matter in the perturbed Sachdev-Ye-Kitaev (SYK) model~\cite{Sachdev_93,micklitz2019non}.

In the spirit of introducing random-matrix ensembles to capture the salient aspects of disorder interacting many-body models, the so-called Rosenzweig-Porter (RP) random-matrix ensemble~\cite{RP} has been introduced as an analytically tractable model having two transitions~\cite{Kravtsov_NJP2015}: the Anderson transition between localized and extended phases, and the other one~--~between ergodic and non-ergodic extended (fractal) ones, see Fig.~\ref{fig:Picture}~(a).
This model is a simple generalization of Gaussian random ensembles, where the off-diagonal matrix elements are scaled down with system size.

Further developments of RP-like models towards more realistic systems found genuine multifractal phases~\cite{LN-RP_RRG,LN-RP_WE,BirTar_Levy-RP} and anomalously slow dynamics~\cite{Monthus,LN-RP_dyn}, realized in a RP model with fat-tail distributed off-diagonals, as well as the effective RP description of non-ergodic phases of matter in disordered many-body systems~\cite{faoro2019non,Tarzia_2020}, Floquet models~\cite{Floquet_MF,Buijsman2022circular}, SYK model~\cite{micklitz2019non}, and graph structures~\cite{LN-RP_RRG,LN-RP_dyn,Bera2018return,DeTomasi2019subdiffusion}.

This work is aimed to inspect the stability of NEE phases in non-Hermitian systems. We focus on the RP model by introducing a non-Hermitian generalization of it. We compute, both analytically and numerically, the phase diagram of our model and show that the ergodic and localized phases are stable with respect to non-Hermiticity.  However, the presence of the intermediate fractal phase depends on the choice of diagonal potential. Indeed, in the case of purely real or imaginary diagonal potential, the fractal phase is intact to any non-Hermiticity (like the complex vector-potential or other non-Hermitian kinetic contributions), and therefore, the phase diagram is the same as the Hermitian RP model, see Fig.~\ref{fig:Picture}~(a). However, the fractal phase disappears for generic complex {\it diagonal} terms, becoming localized. Importantly, these generic complex diagonal terms might mimic experimentally relevant situations, such as random gain and loss contributions. Thus, this gain-and-loss potential gives a possible way to localize the system, unlike the Hatano-Nelson model~\cite{Hatano_96}, in which non-Hermiticity favors delocalization. \rev{The further steps in consideration of more realistic many-body localization models in this gain-loss paradigm are under active investigation~\cite{DeTomasi2022nonHerm_MBL}.}

Panels (a) and (b) of Fig.~\ref{fig:Picture} show the phase diagram for the two aforementioned cases, respectively.  In particular, in Fig.~\ref{fig:Picture}, $\gamma$ is the disorder parameter for the RP model, which controls the level of ergodicity/localization in the system: large $\gamma$ correspond to Anderson insulator, while small values lead to ergodicity in the sense of Gaussian random matrices or Ginibre ensemble (depending on the symmetry). $D$ is the fractal dimension of the wave function. For $D=1$, the system is ergodic, for $D=0$~--~localized, while for $0<D<1$ it has fractal eigenstates. In the final part of our work, we further generalize our model, by interpolating between two cases, from real/imaginary to full complex diagonal potential and study its phase diagram.

The work is organized as follow. In Sec.~\ref{Sec:Model}, we define the model and the probes used to characterize the three phases. In Sec.~\ref{Sec:result} we show the results of the work. The numerical investigation is presented in Sec.~\ref{Sec:result_N}, and the analytical proof in Sec.~\ref{Sec:result_A}. Finally, Sec.~\ref{Sec:conclusion} contains concluding remarks and outlooks.

\section{Model and Methods}\label{Sec:Model}

We consider a generic Ginibre random-matrix ensemble of size $N$, with \rev{Gaussian} independent and identically distributed (i.i.d.) random  elements~\cite{Ginibre_ens}. The off-diagonal elements are rescaled down by the $N$-dependent factor $N^{-\gamma/2}$
\begin{equation}\label{eq:RP_model}
H_{mn} = \zeta_n \delta_{mn} + M_{mn} N^{-\gamma/2} \ ,
\end{equation}
where
\begin{equation}
\overline{\zeta_n} = \overline{M_{mn}} = 0 \ ,
\end{equation}
\begin{equation}
\overline{|\zeta_n|^2} = 1, \quad \overline{|M_{mn}|^2}=\lambda \ .
\end{equation}
The line over the symbols indicates the disorder average.
\revt{The parameter $\lambda$ is relevant only at the Anderson and ergodic transition as it does not affect the scaling of the inverse participation ratio, see, e.g., ~\cite{RP_R(t)_2018}.}

We distinguish the cases of real ($\zeta_n=\ep_n$) and generically complex diagonal matrix elements
\be\label{eq:RP_diag}
\zeta_n = \ep_n + i \nu_n
\ee
as well as we decompose its elements, in Hermitian $h_{mn} = h_{nm}^*$ and anti-Hermitian $a_{mn} = -a_{nm}^*$ parts
\be\label{eq:RP_off-diag}
M_{mn} = h_{mn} + a_{mn} \ .
\ee
In the case of Hermitian $H$, $a_{nm}=0$ and $\nu=0$, the model in Eq.~\ref{eq:RP_model} recovers the usual Hermitian RP model~\cite{RP,Kravtsov_NJP2015}.

In the following two Sections~\ref{Sec:Model_lev} and~\ref{Sec:Model_multi}, we introduce the main probes to distinguish the two phases.

\subsection{Level spacing analysis}\label{Sec:Model_lev}

A powerful probe to distinguish a delocalized phase from a localized one is the statistic of eigenlevel spacing. Here, we focus only on short-range correlations, meaning that we consider only the "gap" statistics of only near\revt{by} energy levels.

In the Hermitian systems, the short range correlation are capture by the so called $r$-gap ratio
\begin{equation}
\label{eq:r_statistic}
    r_n^{R} = \frac{\min{(\Delta_{n+1}, \Delta_n)}}{\max{(\Delta_{n+1}, \Delta_n)}},
\end{equation}
where $\Delta_n = E_{n+1}-E_n$ is the gap between two adjacent energy levels. In the case of an ergodic time-reversal symmetric system $\overline{r_n^{R}}\simeq 0.5307$~\cite{Atas2013distribution,oganesyan2007localization}, which is the same value that for a Hermitian random-matrix. Instead, for localized systems $\overline{r_n^{R}}= 2\log{2}-1 \simeq 0.386$, meaning that the gaps $\{\Delta_n\}$ are Poisson distributed.

In the general non-Hermitian case, $r$-gap statistic in Eq.~\ref{eq:r_statistic} needs to be modified since the spectrum is complex. In Ref.~\onlinecite{Pedro_2020_complex} the gap ratio has been generalized to the non-Hermitian case, by introducing
\begin{equation}
\label{eq:r_statistic_complex}
r_n^{C} = \frac{Z_n^{NN}- Z_n}{Z_n^{NNN} -Z_n},
\end{equation}
where $\{Z_n\}$ is the spectrum of the system and $Z_n^{NN}$ and $Z_n^{NNN}$ are the nearest-neighbor (NN) and the next-to-nearest-neighbor (NNN) of $Z_n$ with respects the \rev{Euclidean distance in the complex $\mathbb{C}$-plane}, respectively.
In general, $r_n^{C} =r_n e^{i\theta_n}\in \mathbb{C}$ and we can analyze $\{r_n\}$ and $\{\theta_n\}$, separately. For Ginibre random-matrix $\overline{-\cos{\theta_n}}\approx 0.229$ and $\overline{r} \approx 0.738$~\cite{Pedro_2020_complex}. Instead, for a localized system, meaning that $\{Z_n\}$ are uncorrelated, we have $\overline{-\cos{\theta_n}}= 0$ and $\overline{r} = 2/3$.

Notice that in the Hermitian problem, the $r$-gap statistics is only able to detect the Anderson transition at $\gamma =2$. This is a special feature of the fractal phase of the RP model, due to the emergence of ergodic energy mini-bands which are ergodic and host the fractal states~\cite{Kravtsov_NJP2015}.

\subsection{Multifractal dimension $D_q$}\label{Sec:Model_multi}

The fractal dimensions $D_q$ quantified the spread of a wave-functions and, it is defined through the inverse participation ratio ($IPR_q$)
\begin{equation}
IPR_q = \sum_{m} |\langle m | R_n \rangle \langle L_n | m \rangle|^{q} \quad q>1/2,
\end{equation}
where $|L_n\rangle$ and $|R_n \rangle$ are the left and the right eigenvector of $H$ with eigenvalue $Z_n$~\footnote{\rev{Here we do not consider other correlations of the left and right eigenvectors known to be non-trivial already for Ginibre~\cite{Mehlig1998Ginibre} and Girko~\cite{Mehlig2000Girko} ensembles.}}. For the Hermitian case, $|L_n\rangle=|R_n \rangle$,  $IPR_q$, defined above, coincides with the standard definition~\cite{Evers2008Anderson}.
The multifractal dimensions are defined as the exponents in the scaling of $IPR_q$ with the matrix size $N$:
\begin{equation}
    IPR_q \sim N^{(1-q)D_q}.
\end{equation}
The ergodic phase is characterized by $D_q=1$, localized~--~by $D_q=0$, while the multifractal phase is given by fractional $0<D_q<1$, being a non-trivial function of $q$.

In the Hermitian RP model, the ergodic ($D_q=1$) and localized ($D_q=0$) phases appear at $\gamma<1$ and $\gamma>2$, respectively, see Fig.~\ref{fig:Fig1}(a). In the intermediate phase, $1<\gamma<2$, the fractal dimension smoothly and linearly interpolates between the above values $D_q=2-\gamma$, see Fig.~\ref{fig:Picture}~(a). Furthermore, one should mention that in the RP model the $D_q$ is independent of $q$, meaning that the phase is \textit{fractal}, but not multifractal. In other words, the fractal states emerging at $1<\gamma<2$ might be understood as an ergodic one living on a manifold of $\sim N^{2-\gamma}$ sites, which are close in their diagonal energies. Such manifolds form fractals~\cite{Kravtsov_NJP2015} in space and compact spectral minibands~\cite{RP_R(t)_2018} in the energy domain, both tiling the entire corresponding spaces.

\section{Results}\label{Sec:result}
In this section, we present our results, first analytical considerations, followed by numerical ones.

\subsection{Analytical Results}\label{Sec:result_A}

\subsubsection{Hermitian RP model}\label{Sec:Hermitian}
We start our discussion by reviewing the Hermitian RP model.
In the Hermitian case, $\nu_n = 0$ and $a_{mn}=0$ in Eqs.~\eqref{eq:RP_diag},~\eqref{eq:RP_off-diag}, and as we discussed in Sec.~\ref{Sec:Model}, the model is  known to have an ergodic/GOE phase ($\gamma<1$), an non-ergodic extended or fractal phase ($1<\gamma<2$), with the fractal dimension $D = 2-\gamma$, and the Anderson localized phase ($\gamma>2$), with the single-site localization~\cite{Kravtsov_NJP2015,RP_R(t)_2018}.

This phase diagram can be found by using the so-called cavity method~\cite{Biroli_RP,Monthus,BogomolnyRP2018}~\footnote{\rev{Note that the phase diagram of the Hermitian Gaussian Rosenzweig-Porter ensemble has been also studied using many other methods. Besides the cavity method, the most significant progress in the analytical studies was achieved by the Dyson Brownian motion technique, which show both the Anderson transition~\cite{Pandey} and the ergodic one~\cite{Biroli_RP}, contour integrals~\cite{BrezHik} and Itzyskon-Zuber formula for the spectral form factor~\cite{ShapiroKunz,Kravtsov_NJP2015}.}}, which connects the Green's function of the system with the Green's function of the same system, where one site is removed, and solved self-consistently. The Green's function is defined as
\be\label{eq:G_def}
G(E+i\delta) = (E + i\delta - H)^{-1} = \sum_n \frac{\Rk{n}\Lb{n}}{E+i\delta - Z_n}\ ,
\ee
written using the spectral decomposition of $H=\sum_n Z_n |R_n\rangle \langle L_n|$, with $Z_n = E_n+i\eta_n$ and
$\LRbk{n}{m}= \delta_{mn}$.

For the Hermitian case, we have $\Lk{n}=\Rk{n}$ and $\eta_n = 0$. In the definition of $G(E)$ in Eq.~\eqref{eq:G_def} as been introduced an infinitesimal regulator $\delta\to 0^+$, to avoid the poles of the Green's function in the real axis.

The cavity equation takes the form:
\be\label{eq:G_cavity_Herm}
G_{ii}(E+i\delta) = \lrb{E+i\delta - \ep_i - \sum_{j,k\ne i} \frac{h_{ij} G_{jk}^{(i)}(E+i\delta) h_{ki}}{N^{\gamma}}}^{-1} \ ,
\ee
where $G^{(i)}(E+i\delta)$ is the Green's function of the problem with removed $i$th row and $i$th column.

We can distinguish two main contributions in
\be
S = \sum_{j,k\ne i} h_{ij} G_{jk}^{(i)}(E+i\delta) h_{ki},
\ee
the diagonal and the off-diagonal
\be
S= S_{diag} + S_{off},
\ee
where
\be
S_{diag} = \sum_{j} |h_{ij}|^2 G_{jj}^{(i)}(E+i\delta) \ .
\ee
$S_{diag}$ is self-averaging and its mean
\be
\overline{S_{diag}} = \tr\lrb{G^{(i)}(E+i\delta)} = N \lrb{\zeta(E)+i\pi\rho(E)},
\ee
and $\rho(E)$ is the density of states. The variance of of real and imaginary parts of $S_{diag}$
\be
\overline{\delta \lrp{\Re/\Im S_{diag}}^2} = {2\sum_j\lrb{\Re/\Im G^{(i)}(E+i\delta)}^2} \simeq O(N) \ ,
\ee
As a result, we obtain the self-averaging property, $\overline{\delta \lrp{\Re/\Im S_{diag}}^2}/\overline{S_{diag}}^2\rightarrow 0$ and, $S_{diag}\simeq \overline{S_{diag}}$.

Instead, the second contribution
\be
S_{off} = \sum_{j,k\ne i\atop j\ne k} h_{ij} G_{jk}^{(i)}(E+i\delta) h_{ki}
\ee
has zero mean, because $\overline{h_{i<j}}=0$, and the variance of real and imaginary parts
\be
\begin{split}
\overline{\lrp{\Re/\Im S_{off}}^2} =  2\sum_{j,k\ne i\atop j\ne k} \overline{h_{ij}^2 \lrp{\Re/\Im G_{jk}^{(i)}(E+i\delta)}^2 h_{ki}^2} \\ =   2\lambda^2 \sum_{j,k\ne i\atop j\ne k} \overline{\lrp{\Re/\Im G_{jk}^{(i)}(E+i\delta)}^2} \sim O(N^2) \ .
\end{split}
\ee
Notice that for simplicity we considered real symmetric $h_{ij} = h_{ji} = h_{ij}^*$ leading to symmetric $G_{jk} = G_{kj}$.

Usually, the Green's function is dominated by the diagonal elements, therefore the standard deviation of the second contribution is much smaller (but not parametrically smaller) than the mean of the first contribution.

All this leads to the standard formula (self-)averaged over the off-diagonal elements
\be\label{eq:G_aver_Herm}
\overline{G_{ii}(E+i\delta)}_{h} = \lrb{E+i\delta - \ep_i - N^{1-\gamma}\lrp{\sigma(E)+i\pi \rho(E)}}^{-1} \ .
\ee
The level broadening defined as
\be
\Gamma_i = \pi\rho(E)\sum_{j}\abs{H_{ij}}^2 \sim \rho(E)N^{1-\gamma}
\ee
is in agreement with the Fermi's Golden rule result, while
\revt{the real part of the self-energy} $N^{1-\gamma}\sigma(E)$ is an unimportant (nearly constant) term, which is small for $\gamma>1$ and can be absorbed by the energy shift of the diagonal matrix elements.
For $\gamma<1$ both the shift and the level broadening are large compared to the diagonal energies $\ep_i$, therefore neither of them can be absorbed by the energy shift\rev{. These large amplitudes of the self-energy values diminish any dependence of $\mean{G_{ii}(E+i\delta)}_{h}$ on the coordinate $i$ via $\ep_i$}. As a result, $\gamma<1$ corresponds to ergodicity.

On the other hand, in a fractal phase, $1<\gamma<2$, the broadening $\Delta\ll\Gamma = N^{1-\gamma} \pi\rho(E)\ll O(1)$ is simultaneously large compared to the mean level spacing $\Delta\sim 1/N$ and small compared to an entire spectral bandwidth $\sim O(1)$.
Thus, \revt{from Eq.~\eqref{eq:G_aver_Herm} one can immediately see that the number of sites $i$, where the wave function is of the same order as at the maximum is given by
\be
|\ep_i-E|\lesssim \Gamma
\ee
This confirms the fact the the wave-function structure is fractal (but not multifractal) and the fractal dimension $D_{q>1}=D$ is given by the support set $N^D\sim \Gamma/\Delta$. The scaling of} $\Gamma$ leads to a finite fractal dimension $D = 2-\gamma$

\rev{Formally one can see that the above results seem to be applicable to a non-Hermitian cases.
However, strictly speaking, as mentioned in Sec.~2.3 of~\cite{Metz2019spectral}, the framework of a Green's function (or a resolvent) may fail to describe the continuous (bulk) part of the spectrum in the non-Hermitian case. Therefore in the next section we take another route.}

\subsubsection{Non-Hermitian RP Model}
\revt{In this section we consider a method, similar to the one, considered in Sec.~\ref{Sec:Hermitian}, applicable (unlike the previous one)} to the non-Hermitian case in Eq.~\ref{eq:RP_model}. \revt{Indeed, in} the non-Hermitian case, both the diagonal~\eqref{eq:RP_diag} and off-diagonal~\eqref{eq:RP_off-diag} elements of the matrix~\eqref{eq:RP_model} can be complex and non-Hermitian.
\rev{This leads, in particular, to the fact that the Green's function framework cannot describe} the continuous part of the spectrum~\eqref{eq:G_def}~\cite{Metz2019spectral}.

In order to deal with the non-Hermitian matrices properly, one should consider the Hermitization of the problem~\cite{FEINBERG_1997,FEINBERG1_1997,Feinberg_2001,Metz2019spectral} (see, e.g., Sec. 2.4 in~\cite{Metz2019spectral} for review).
Instead of $N\times N$ matrix of Hamiltonian $H$ shifted by a complex energy $z = E +i\eta$, with generic real $E$ and imaginary $\eta$ parts, we consider the following
$2N\times 2N$ Hermitian matrix
\be
B(z)= B^\dagger(z) = \left(
      \begin{array}{cc}
        0 & H - z \\
        H^\dagger - z^* & 0 \\
      \end{array}
    \right)
\ee
and determine the analog of the Green's function as the inverse of $B(z)$ with an infinitesimal regularizer $\delta\to 0^{+}$

\be\label{eq:G_def-2}
\G(z,\delta) = \lrb{B(z) + i\delta}^{-1} = 
\left(
      \begin{array}{cc}
        -i\delta\cdot X & X(H - z) \\
        (H^\dagger - z^*) X & -i\delta\cdot \tilde{X} \\
      \end{array}
    \right)
\ ,
\ee
where we used a standard block-matrix inversion
and introduced two matrices
\be
\begin{split}
X &= X^\dagger = \lrb{\delta^2 + (H - z)(H^\dagger - z^*)}^{-1} \ , \\
\tilde X &= \tilde X^\dagger = \lrb{\delta^2 + (H^\dagger - z^*)(H - z)}^{-1} \ .
\end{split}
\ee
which are similar to each other by the transformations
\be\label{eq:tildeX_via_X}
\begin{split}
(H-z) \tilde X & = X (H-z), \\
\tilde X (H^\dagger-z^*) & = (H^\dagger-z^*) X  \ .
\end{split}
\ee

Importantly, the generalized Green's function in Eq.~\eqref{eq:G_def-2} has direct access to the eigenvectors $\Lk{n}$, $\Rk{n}$ and eigenvalues $Z_n = E_n + i\eta_n$ of the non-Hermitian problem. Indeed, one can show that~\cite{Metz2019spectral} at $Z=Z_n$
\be\label{eq:B_mat_props}
B(Z_n) \ket{b_n^\pm} = 0 \text{, with } \ket{b_n^\pm} = \lrp{\pm \Lk{n} \atop \Rk{n}},
\ee
and therefore
\be
i \delta\cdot \G(Z_n,\delta) =
2
\begin{pmatrix}
\Lk{n}\Lb{n} & 0 \\
0 & \Rk{n}\Rb{n}
\end{pmatrix}
 + O(\delta)
\ .
\ee
\rev{Note that the Hermitization technique doubles the number of eigenstates and therefore may affect the eigenvalue statistics. Therefore here we use this technique only for the description of eigenstate (not eigenvalue) statistics, which is free from this drawback~\footnote{\rev{
In the numerical simulations below we diagonalize directly non-Hermitian matrices without using this Hermitization technique.}}}

Following Ref.~\onlinecite{Metz2019spectral}, one can write the analogue of the cavity equation for the non-Hermitian case
\be\label{eq:G_cavity}
\G_{ii}(z,\delta) = \lrb{Z-i\delta - \z_i - N^{-\gamma}\sum_{j,k\ne i} \h_{ij} \G_{jk}^{(i)}(z,\delta) \h_{ki}}^{-1} \ ,
\ee
where
\be\label{eq:Green_func_non-Herm}
\G_{ij}(z,\delta)= \left(
         \begin{array}{cc}
           -i \delta\cdot X_{ij}  & \lrb{X(H-z)}_{ij} \\
           \lrb{(H^\dagger-z^*)X}_{ij} & -i\delta\cdot \tilde X_{ij} \\
         \end{array}
       \right) \ ,
\ee
\be
\z_i = \left(
         \begin{array}{cc}
           0 & z_i \\
           z_i^* & 0 \\
         \end{array}
       \right) \ , \quad
Z = \left(
         \begin{array}{cc}
           0 & z \\
           z^* & 0 \\
         \end{array}
       \right) \ , \quad
\h_{ij} = \left(
         \begin{array}{cc}
           0 & M_{ij} \\
           M_{ji}^* & 0 \\
         \end{array}
       \right) \ ,
\ee
and all the matrices ($B(z)$, $\G(z,\delta)$ and others) are rewritten in the basis which is reordered as
\be
\lrp{1,\ldots,N,N+1,\ldots,2N} \to
\lrp{1,N+1,2,N+2\ldots,N,2N} \ .
\ee
The convenience of this relabeling is to
allow one to have the above $2\times 2$ blocks like $\z_i$.

Finally, similarly to the Hermitian case, we can distinguish two contributions to the self-energy
\be
\s(z,\delta) = \sum_{j,k\ne i}\h_{ij} \G_{jk}^{(i)}(z,\delta) \h_{ki} = \s_{diag} + \s_{off} \ .
\ee
The diagonal contribution
\be
\s_{diag}(z,\delta) =
\ee
\be
=\sum_{j\ne i}\left(
         \begin{array}{cc}
           -i \delta\cdot \tilde X_{jj}|M_{ij}|^2  & \lrb{(H^\dagger-z^*)X}_{jj}M_{ij}M_{ji} \\
           \lrb{X(H-z)}_{jj}M_{ji}^*M_{ij}^* & -i\delta\cdot X_{jj}|M_{ji}|^2 \\
         \end{array}
       \right),
\ee
Its average over matrix ensemble
\be
\overline{\s_{diag}(z,\delta)} = - i\delta \tr X\left(
         \begin{array}{cc}
           1 & 0 \\
           0 & 1 \\
         \end{array}
       \right)\equiv -i N  \delta \bar X,
\ee
where we have used the fact that $X$ and $\tilde X$ are similar~\eqref{eq:tildeX_via_X} and Hermitian, therefore have the same real spectra. In particular,
\be
\tr X(z,\delta) = \tr\tilde X(z,\delta) \ .
\ee

Neglecting the off-diagonal part for the same reasons as in the Hermitian case and using Eq.~\eqref{eq:G_def-2}, one obtains for $\G_{ii}$ the following self-consistent equation
\begin{multline}\label{eq:Green_funct_ii_non-Herm}
\G_{ii}(z,\delta) =
\left(
      \begin{array}{cc}
        -i\delta\cdot X_{ii} & \lrb{X(H - z)}_{ii} \\
        \lrb{(H^\dagger - z^*)X}_{ii} & -i\delta\cdot \tilde{X}_{ii} \\
      \end{array}
    \right) = \\
=\left(
      \begin{array}{cc}
        i\delta\cdot \lrp{1 + N^{1-\gamma} \bar X} & z - Z_i \\
        z^* - Z_i^* & i\delta\cdot \lrp{1 + N^{1-\gamma} \bar X} \\
      \end{array}
    \right)^{-1} = \\=
R\left(
      \begin{array}{cc}
        -i\delta\cdot \lrp{1 + N^{1-\gamma} \bar X} & z - Z_i \\
        z^* - Z_i^* & -i\delta\cdot \lrp{1 + N^{1-\gamma} \bar X} \\
      \end{array}
    \right)
    \ ,
\end{multline}
where
\be\label{eq:Denom_Gii}
R = \frac{1}{|z-Z_i|^2+\delta^2 \lrp{1 + N^{1-\gamma} \bar X}^2},
\ee
and
\be\label{eq:<X>}
\bar X = \frac1{N}\sum_n {\overline{X_{nn}}}.
\ee
\begin{figure}[ht!]
\label{fig:Fig1}
    \includegraphics[width=1.\columnwidth]{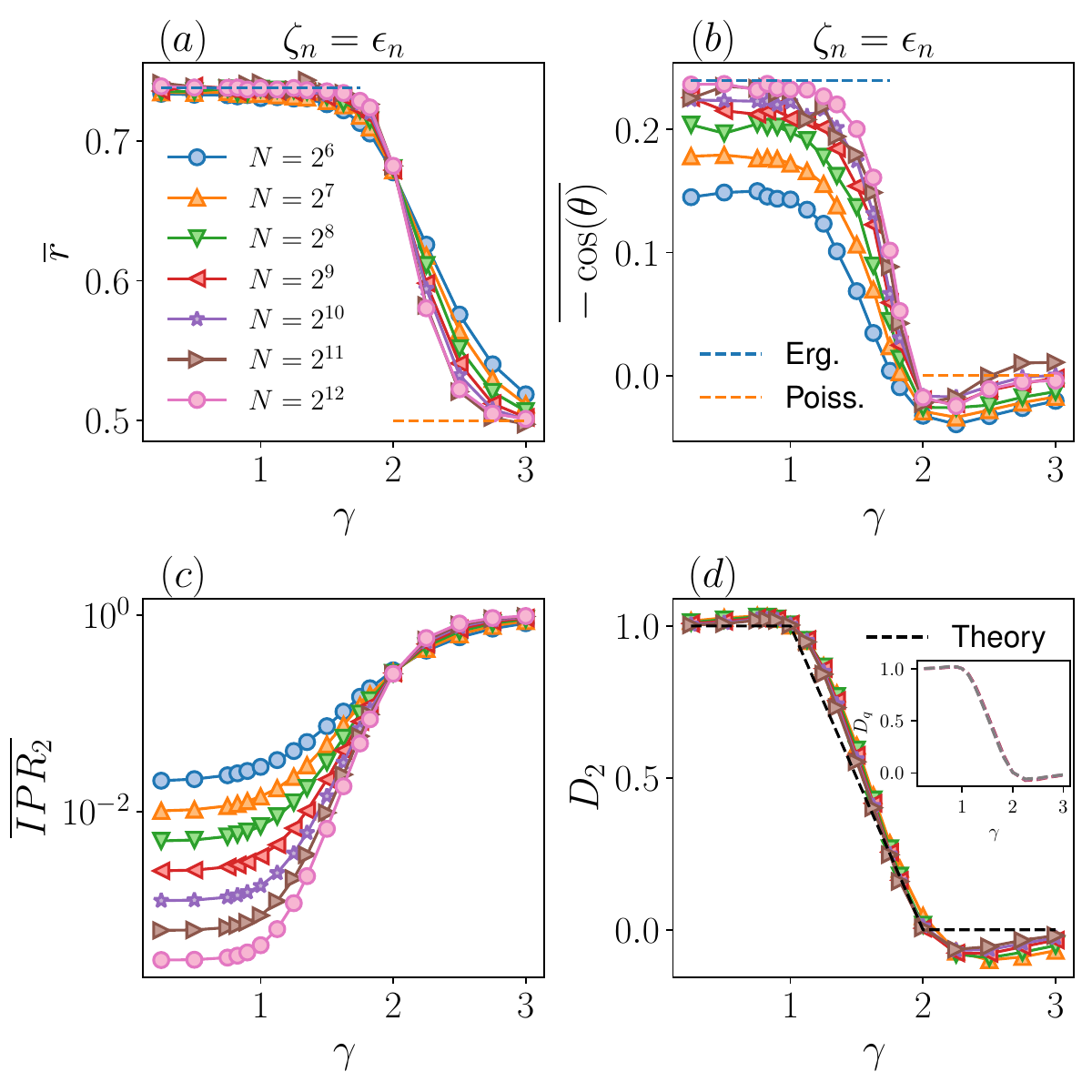}
    \caption{\textbf{Phase diagram of the Rosenzweig-Porter model in the case of purely real/imaginary potential: $z_n = \epsilon_n$.} (a),(b)~radial and angular components of the complex gap ratio, Eq.~\ref{eq:r_statistic_complex}, respectively.
    (c)~inverse participation ratio ($IPR_2$). (d)~fractal dimension $D_q$ extracted from $IPR_q$ with $q=2$ for different system sizes \rev{as $D_2(N) = \ln[IPR_2(N)/IPR_2(N')]/\ln[N'/N]$} (inset: for $q\in \{1.5,1.75,2,2.5,3\}$). The dashed lines in (a), (b) are the ergodic (Erg.) and Poisson values. The black dashed line in (d) is the theory prediction ($N\rightarrow \infty$).
        \rev{Here and further in the next figures, all the measures are averaged over the entire spectrum.
        The fractal dimensions in panel (d) show some overshooting close to the transition points $\gamma=1$ and $2$ due to the finite-size effects, given by the extraction from two relatively modest system sizes.}
    }
\end{figure}
Here we introduce the complex analogue of the broadening $\Gamma_n = \delta\cdot X_{nn} N^{1-\gamma} \equiv \Delta N^{D}$ and determine the fractal dimension using the mean level spacing $\Delta$.
The self-consistency equation for $\bar \Gamma$
\rev{ can be derived by the substitution of the definition of the broadening $\Gamma_n$ and the diagonal part $X_{nn}$ of the generalized Green's function~\eqref{eq:Green_func_non-Herm} together with the expressions~\eqref{eq:Green_funct_ii_non-Herm},~\eqref{eq:Denom_Gii} to Eq.~\eqref{eq:<X>}.
Eventually it}
takes the form
\be\label{eq:Gamma_expr}
\bar \Gamma = N^{-\gamma} \sum_n \frac{\delta +\bar \Gamma}{|z - Z_n|^2 +\lrp{\delta +\bar \Gamma}^2}
\ee
For $\gamma<1$, it is obvious that $\bar \Gamma \gg 1$ and thus one can neglect $z-Z_n$ in the expression for $\Gamma_{nn}$ in the sum~\eqref{eq:<X>}. As a result, for $\gamma<1$
\be
\bar \Gamma \simeq \frac{N^{1-\gamma}}{\delta + \bar \Gamma} \lra
\bar \Gamma \sim N^{(1-\gamma)/2}\gg 1 \ ,
\ee
leading to an ergodic behavior, like to the Hermitian case.
\begin{figure}[ht!]
\label{fig:Fig2}
    \includegraphics[width=1.\columnwidth]{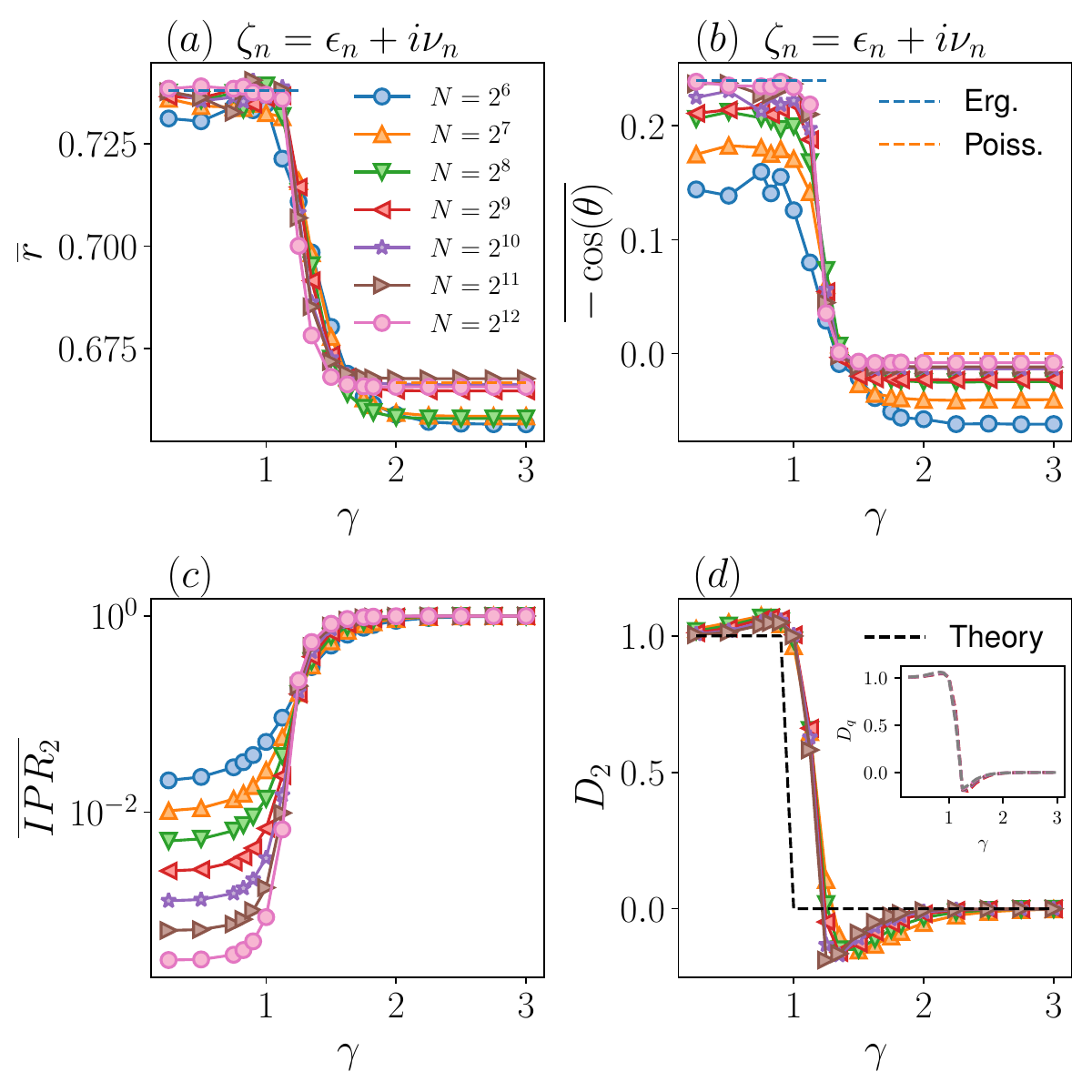}
    \caption{\textbf{Phase diagram of the RP-model in the case of complex potential: $z_n = \epsilon_n + i\nu_n$.} (a),~(b)~complex gap ratio defined in  Eq.~\ref{eq:r_statistic_complex} and below it.
    (c)~inverse participation ratio ($IPR_2$). (d)~fractal dimension $D_2$ for different system sizes \rev{as $D_2(N) = \ln[IPR_2(N)/IPR_2(N')]/\ln[N'/N]$}. (inset)~$D_q$ for $q\in \{1.5,1.75,2,2.5,3\}$). The dashed lines in (a), (b) represent the ergodic (Erg.) and Poisson values. The black dashed line in (d) gives the theoretical
    prediction Eq.~\ref{eq:D_q_general}.
    }
\end{figure}

For $1<\gamma<2$, one should consider two different cases:
\begin{enumerate}
    \item If $Z_n$ is real (or imaginary), then one can replace the summation in Eq.~\eqref{eq:<X>} by $1$d integration using $\Delta Z_n \simeq 1/N$
    \be\label{eq:Y_real}
    \begin{split}
    \frac{\bar \Gamma}{N^{1-\gamma}} \simeq & \int_{-1}^1 \frac{\delta+ \bar \Gamma}{|z-Z_n|^2+\lrp{\delta+ \bar \Gamma}^2}dZ_n = \\
    = & 2 {\rm arccot}\lrp{\delta+ \bar \Gamma} \simeq O(1) \ ,
    \end{split}
    \ee
    giving the usual fractal dimension\revt{, see dashed line in Fig.~\ref{fig:Fig1}(d)}
    \be
    D_q = \caseS{
    \begin{array}{ll}
    1, & \gamma<1\\
    2-\gamma, & 1<\gamma<2\\
    0, & \gamma>2
    \end{array}
    } \ ,
\ee
    as in the Hermitian case~\cite{Kravtsov_NJP2015}.
\begin{figure}[t!]
\label{fig:Fig3}
    \includegraphics[width=1.\columnwidth]{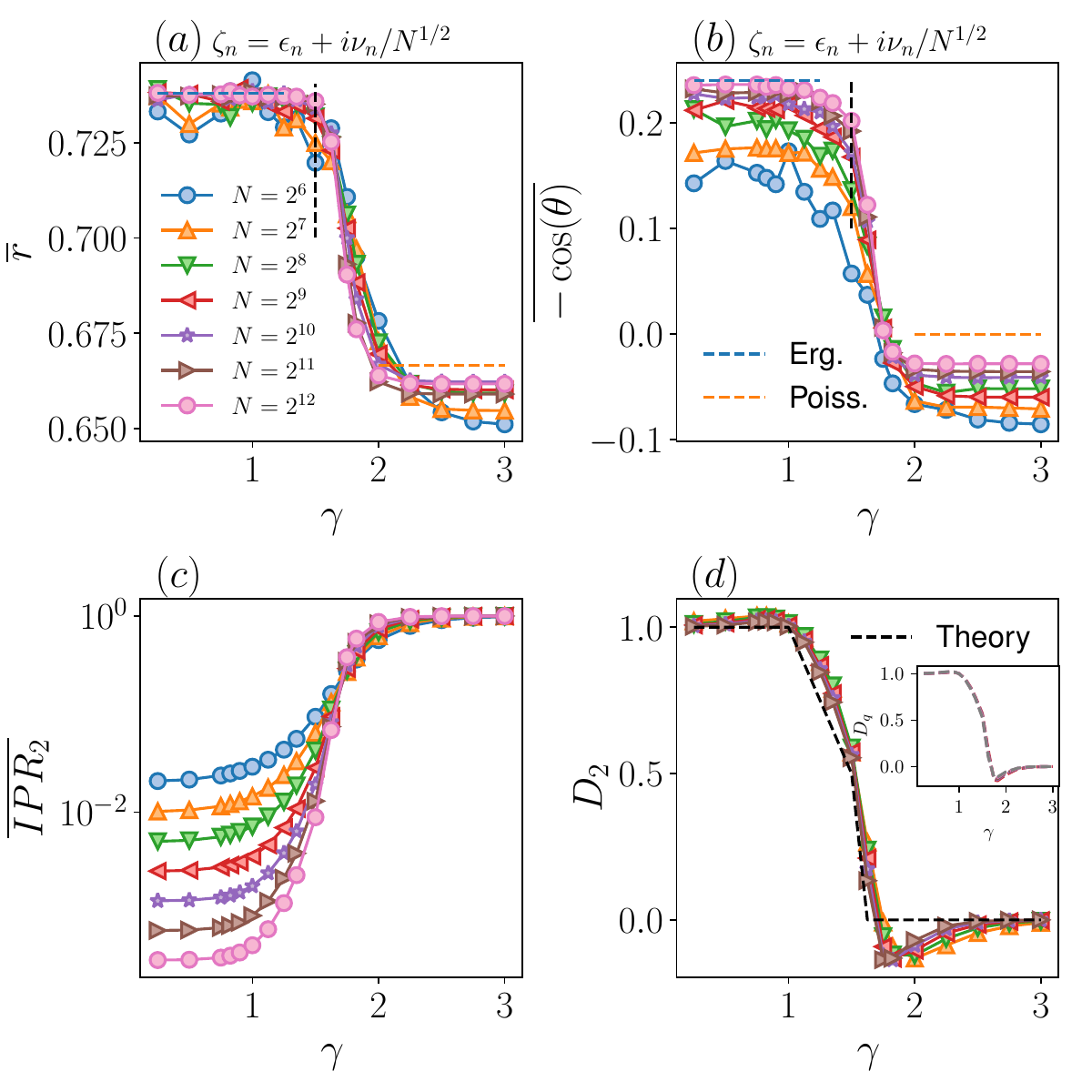}
    \caption{\textbf{Phase diagram of the RP-model in the case of complex potential with rescaled imaginary part: $z_n = \epsilon_n + i\nu_n/N^{1/2}$.} (a),(b)~complex gap ratio defined in  Eq.~\ref{eq:r_statistic_complex}.
    (c)~inverse participation ratio ($IPR_2$) for several $N$. (d)~fractal dimension $D_2$ \rev{extracted from the} different system sizes \rev{as $D_2(N) = \ln[IPR_2(N)/IPR_2(N')]/\ln[N'/N]$}. (inset)~$D_q$ for $q\in \{1.5,1.75,2,2.5,3\}$). The horizontal dashed lines in (a), (b) represent the ergodic (Erg.) and Poisson values, while the vertical black dashed lines sign the value of the critical point $\gamma=1+c=3/2$ The black dashed line in (d) gives the theoretical prediction in Eq.~\ref{eq:D_q_general_c}.
    }
\end{figure}

    \item In the more general case,  $Z_n$ is complex, $Z_n = \epsilon_n + i \nu_n$, one can replace the summation in Eq.~\eqref{eq:<X>} by a $2$d integration.
    As soon as the number of point $Z_n$ in the complex plane is still given by $N$, for each of them the real $\epsilon_n$ and imaginary $\nu_n$ parts of $Z_n$ have the average discrete steps given by $\Delta \epsilon_n \Delta \nu_n \simeq 1/N$ and they are of the same order  $\Delta \epsilon_n = \Delta \nu_n \simeq 1/N^{1/2}$. This gives (for $z=0$)
    \begin{multline}\label{eq:Y_complex}
    \frac{\bar \Gamma}{N^{1-\gamma}} \simeq \iint_{-1}^1 \frac{\delta+ \bar \Gamma}{|\epsilon_n + i \nu_n-z|^2+\lrp{\delta+ \bar \Gamma}^2}d\epsilon_n d\nu_n =\\=
    \int_{-1}^1 2 {\rm arccot}\lrb{\sqrt{\lrp{\delta+ \bar \Gamma}^2+\nu_n^2}}\frac{\delta+ \bar \Gamma}{\sqrt{\lrp{\delta+ \bar \Gamma}^2+\nu_n^2}}d\nu_n \\
    \simeq \pi\lrp{\delta+ \bar \Gamma}\ln\lrp{\frac{C}{\delta+ \bar \Gamma}} \ ,
    \end{multline}
    with a certain unimportant constant $C\sim O(1)$. In the limit $\delta\to 0$, one can cancel $\bar \Gamma$ in both sides of the equation and obtain

\be
\ln\lrp{\frac{\bar \Gamma}C} =  -\frac1{\pi}N^{\gamma-1}\gg 1,
\ee
as a result
\be
\bar \Gamma = C e^{-N^{\gamma-1}/\pi}\ll N^{-1} \ .
\ee
In the opposite limit of $N\to \infty$ before $\delta\to 0$, one obtains
\be
\bar \Gamma \simeq \pi N^{1-\gamma} \delta\ln\lrp{\frac{C}{\delta}}
\ee
The smallness of $\bar \Gamma$ leads to the localization $D_q = 0$ for all $\gamma>1$\revt{, see dashed line in Fig.~\ref{fig:Fig2}(d)},
\be \label{eq:D_q_general}
    D_q = \caseS{
    \begin{array}{ll}
    1, & \gamma<1\\
    0, & \gamma>1
    \end{array}
    } \ .
\ee
\end{enumerate}

Summarizing, for $\gamma<1$ or $\gamma>2$ the system is ergodic or localized, respectively. For $1<\gamma<2$ the system hosts a fractal phase with $D_q=2-\gamma$ if $\{Z_n\}$ is purely real or imaginary. Instead, in the general case $Z_n= \epsilon_n+i\nu_n$, for $1<\gamma <2$ the system is localized $D_q=0$.

We can consider a more general case, which interpolate the two cases, in which the diagonal elements of $H$ in Eq.~\ref{eq:RP_model} are given by
\be\label{eq:def_c}
\zeta_n = \epsilon_n + i\nu_n/N^{c}.
\ee
For $c>1$, we recover the first case, while with $c=0$ the second case. Using similar arguments as in Eq.~\ref{eq:Y_complex}, one finds\revt{, see dashed line in Fig.~\ref{fig:Fig3}(d)}
\be\label{eq:D_q_general_c}
    D_q = \caseS{
    \begin{array}{ll}
    \min(1,2-\gamma), & \gamma<1+c\\
    0, & \gamma>1+c
    \end{array}
    } \ .
\ee

\subsubsection{Qualitative understanding of the results}

\begin{figure}[t!]
\label{fig:Cartoon_results}
    \includegraphics[width=0.5\textwidth]{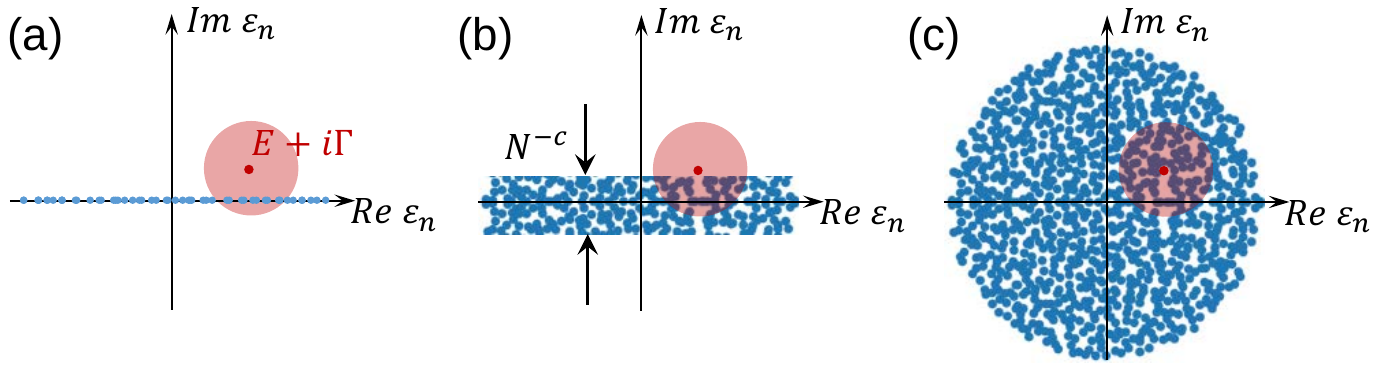}
    \caption{\textbf{Qualitative explanation of the results.}
    The level broadening $\Gamma$ (pink circle) of an energy level $E$ is plotted on top of the distribution of diagonal matrix elements (blue dots) for 3 different cases:
    (a)~real diagonal, (b)~the interpolating regime with $N^{-c}$ imaginary term amplitude in Eq.~\ref{eq:def_c}, (c)~generic complex diagonal.
    }
\end{figure}

All the above results can be straightforwardly understood already at the level of a usual Green's function~\eqref{eq:G_def} (even though formally it is not applicable for the description of the continuous spectral part, see~\cite{Metz2019spectral} for details).

Indeed, in the Hermitian case with $\gamma>1$ the real energy shift $N^{1-\gamma}\sigma(E)$ is not important as it can be absorbed by the energy shift of the real diagonal matrix elements (see Fig.~\ref{fig:Cartoon_results}(a)).
In the non-Hermitian case with purely real or imaginary diagonal elements, the situation is the same (with respect to a rotation in a complex plane) as either the imaginary or real part of the broadening cannot be absorbed by a simple energy shift.

The situation changes drastically, however, as soon as the diagonal term becomes genuinely complex. In this case, one should compare the broadening $\sim N^{1-\gamma}$ with the dimensions of the diagonal element distribution in a complex plane.
Indeed, as soon as $N^{-c}<\Gamma$, i.e. $\gamma<1+c$, the diagonal terms cannot absorb the broadening (along the imaginary axis) and the phase diagram should repeat the Hermitian case.
However, in the opposite limit of $\gamma>1+c$, when the diagonal element distribution ``buries'' both real and imaginary parts of the broadening, the latter becomes a simple energy shift and the situation becomes equivalent to the single-site localized phase, present in the Hermitian case at $\gamma>2$, see Fig.~\ref{fig:Cartoon_results}(b).

\subsection{Numerical Results}\label{Sec:result_N}

Having established the phase diagram of the RP model analytically, we now confirm our results numerically.
\begin{figure}[t!]
\label{fig:Fig4}
    \includegraphics[width=1.\columnwidth]{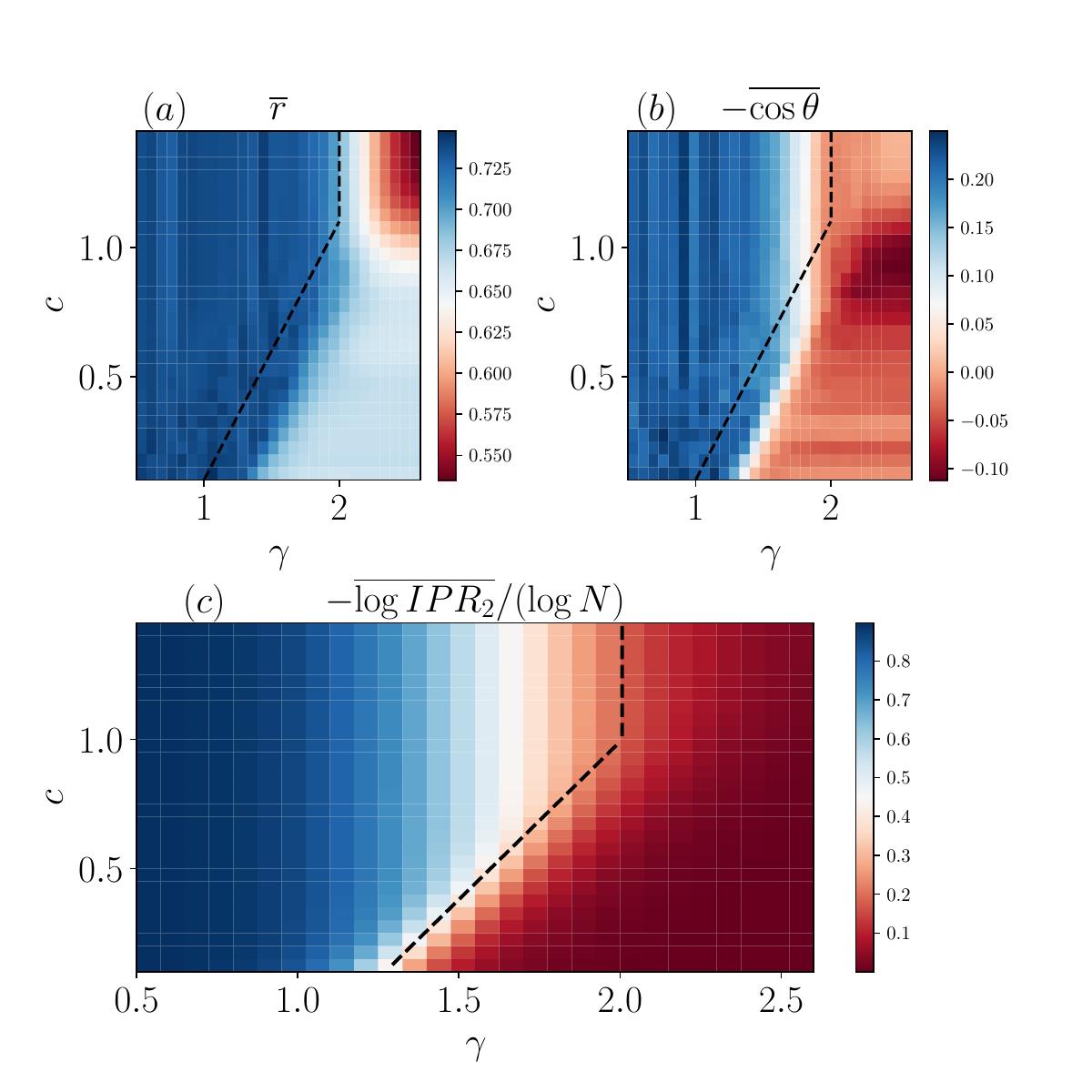}
    \caption{\textbf{Phase diagram of the Rosenzweig-Porter model in the case of complex potential with rescaled imaginary part: $z_n = \epsilon_n + i\nu_n/N^{c}$.} In the $x-$axis, we have the disorder parameter $\gamma$ of the RP-model and in $y-$axis is the $c$ parameter in $z_n = \epsilon_n + i\nu_n/N^{c}$.
    (a),~(b)~The intensity in the figure stands for the complex gap ratio defined in  Eq.~\ref{eq:r_statistic_complex}.
    (c)~Finite size fractal dimension $D_2=-\frac{\overline{\log{IPR_2}}}{\log{N}}$. In all plots $N=2^{10}$.
    The black dashed lines are guides for eyes, indicating the theoretical prediction for the Anderson transition.
    }
\end{figure}
We start with the investigation of the spectrum of the generalized RP model defined in Eq.~\ref{eq:RP_model} in the case of the purely real ($\zeta = \epsilon_n$) diagonal elements. Figure~\ref{fig:Fig1}~(a)-(b) shows the radial ($\overline{r}$) and angular component ($-\overline{\cos(\theta)}$) for the complex gap energy statistic defined in Eq.~\ref{eq:r_statistic_complex} as a function of the tuning parameter $\gamma$ for several $N\in [2^6,2^{12}]$. As one can notice for $\gamma <2$, both $\overline{r}$ and $-\overline{\cos{\theta}}$ reach their ergodic value, meaning that the gap statistic is similar to the one of a Ginibre  random-matrix. Instead, for $\gamma>2$ the $\overline{r}$ and $-\overline{\cos(\theta)}$ tend to the corresponding Poisson value. Notice, that in this case $\overline{r}_{Poisson} = 1/2$, since the spectrum is real in the limit $N\rightarrow \infty$. This is in agreement with an Anderson transition, between delocalized to localized at $\gamma = 2$.

As pointed out in Ref.~\onlinecite{Kravtsov_NJP2015}, the gap energy statistics does not give access to the fractal nature of the wave-function, but it is only \revt{sensitive} to separation between extended and localized states. To detect the fractal phase, we investigate the scaling of $IPR_q$ with $N$ to extract the fractal dimension $D_q$. Figure.~\ref{fig:Fig1}~(c)-(d) shows $IPR_2$ for several $N$ and $D_2$, respectively. $D_2$ has been computed considering the $log-$ derivative, $D_q(N) = -\frac{1}{1-q}\frac{d \log{IPR_q}}{d\log(N)}$. In good agreement with our analytical considerations, we found that $D_2 =1$ for $\gamma <1$, $D_2=2-\gamma$  for $1<\gamma<2$ and, $D_2=0$ for $\gamma>2$. Furthermore, we check the the fractal nature of the intermediate phase, meaning that $D_q$ is $q$-independent, as shown in the inset of Fig.~\ref{fig:Fig1}~(d).

In Figs~\ref{fig:Fig2} and~\ref{fig:Fig3}, we analyze numerically the two remaining cases. First, in Fig.~\ref{fig:Fig2} we demonstrate the case in which the diagonal potential is generically complex ($\zeta_n = \epsilon_n +i\nu_n$, $\mean{\epsilon_n^2}\simeq \mean{\nu_n^2}$). Here the complex gap ratio undergoes the transition around $\gamma=1$ from ergodic Ginibre to Poisson values. Note that due to the complex nature of the diagonal potential, the Poisson value of the absolute value is $\bar r = 2/3$~\cite{Pedro_2020_complex}.
The $IPR_q$ and the corresponding fractal dimension show the corresponding behavior. The fractal dimension experiences a jump close to $\gamma\simeq 1$ from $D=1$ to $0$, implying a direct Anderson transition from the ergodic to the localized phase.

Second, as an intermediate case we consider $\zeta_n = \epsilon_n +i\nu_n/N^{c}$ with $c=1/2$ in~\ref{fig:Fig3}.
In this case, the gap ratio characteristics undergo a transition at $\gamma = 1+c = 1.5$ between ergodic and localized values, while
the fractal dimension first follows the smooth linear curve $D=2-\gamma$ of the Hermitian RP model and then experiences a jump to zero at $\gamma=1+c$.

In both cases, the scaling analysis is in good agreement with the analytical arguments and in particular with the analytical formula for the fractal dimension $D_q$.

Finally, in Fig.~\ref{fig:Fig4} we summarize our results in a density plot, where $x-$axis is $\gamma$, the $y-$axis $c$ in Eq.~\ref{eq:def_c}. The color intensity in Fig.~\ref{fig:Fig4}~(a)-(c) stands for $\overline{r}$, $-\overline{\cos{\theta}}$ and $D_2$ for $N=2^{10}$, respectively.  As one can observe, for $c\approx 0$ the Anderson transition is at $\gamma=1$ and from there opens a fractal phase in $1<\gamma<1+c$, as predicted in Eq.~\ref{eq:D_q_general_c}.

\section{Conclusions}\label{Sec:conclusion}
In this work, we inspect the robustness of fractal states to non-Hermiticity. We generalized the RP model to the non-Hermitian case both in terms of gain and loss and kinetic Hatano-Nelson contributions. The Hermitian RP model is known to host three different phases: an ergodic, fractal, and localized one. We provide both numerical and analytical evidence that the phase diagram of the model depends only on the non-Hermitian properties of the diagonal potential, but not on those of the kinetic (off-diagonal) term.
Indeed, the phase diagram is unchanged compared to the Hermitian case if the diagonal terms are purely real or imaginary. For a generic complex potential, corresponding to gains and losses, the fractal phase disappears, surprisingly, giving the way to a localized one. The diagonal term, whenever is purely real/imaginary or complex, changes the phase diagram of the RP model, while the nature of the off-diagonal terms barely affects it. Finally, we redefine our model by introducing a new tuning parameter, which enables us to interpolate between the above two cases, and to study the full phase diagram.

Our work paves the way for the study of non-ergodic extended phases of matter in non-Hermitian quantum systems. Our generalized RP model could be thought of as an effective model for disordered non-Hermitian many-body systems and could have application in open or monitored systems. Understanding the impact of the nature of the potential in local many-body systems and the generalization of other random-matrix ensembles to the non-Hermitian case, remain objectives for future research~\cite{DeTomasi2022nonHerm_MBL}.

\section{Acknowledgments}
We are grateful to V. E. Kravtsov for stimulating discussion.
I.~M.~K. acknowledges the support
by the European Research Council under the European
Union's Seventh Framework Program Synergy
ERC-2018-SyG HERO-810451.
G.~D.~T. acknowledges the support from the EPiQS Program of the Gordon and Betty Moore Foundation.

\bibliography{Lib}

\end{document}